\documentclass[twocolumn,showpacs,preprintnumbers,amsmath,amssymb,prb,floatfix]{revtex4}

\usepackage{graphicx}
\usepackage{dcolumn}
\usepackage{bm}
\usepackage{array}
\usepackage{amsmath,amssymb}

\newcommand{\be}{\begin{eqnarray}}
\newcommand{\ee}{\end{eqnarray}}

\begin{document}
\title{Gauge fields, quantized fluxes and  monopole confinement of the honeycomb lattice}
\author{ Mahito Kohmoto }
\affiliation{Institute for Solid State Physics, University of Tokyo, 5-1-5 Kashiwanoha, Kashiwa, Chiba 277-8581, Japan\\}
\begin{abstract}
Electron hopping models on the honeycomb lattice are studied. The lattice consists of two triangular  sublattices, and  it is non-Bravais. The dual space has non-trivial topology. 
The gauge fields of Bloch electrons have the $U(1)$ symmetry and thus represent superconducting states in the dual space. Two quantized Abrikosov fluxes exist at the Dirac points and have fluxes $2\pi$ and $-2\pi$, respectively.
We define the non-Abelian $SO(3)$ gauge theory in the extended 3$d$ dual space and it is shown that a monopole and anti-monoplole solution is  stable. The $SO(3)$ gauge group  is broken down to $U(1)$ at the 2$d$ boundary.
The Abrikosov fluxes are related to quantized Hall conductance by the topological expression. Based on this, monopole confinement and deconfinement are discussed in relation to time reversal symmetry and QHE.

The Jahn-Teller effect is briefly discussed.
\end{abstract}


\maketitle
\section{ Introduction} 
There have been much scientific and  technological interests in Carbon compounds. Notable examples are carbon nanotubes and {\em graphene}, a single layer of carbon atoms\cite{novo,zhang,berger}. Integer quantum Hall effect  has been reported in graphene\cite{novo,zhang}. In both materials carbon atoms form the honeycomb lattice.

\begin{figure}[t]
\begin{center}
\includegraphics[width=55mm]{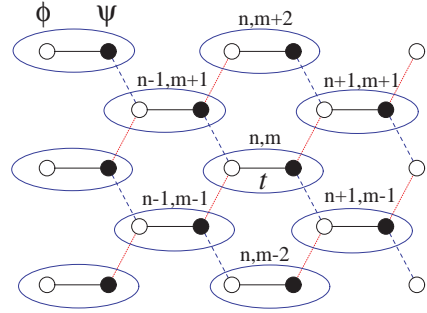}
\caption{(Color online) The honeycomb lattice. Black and white circles are lattice points on sublattice A and B, respectively. Horizontal bonds between them are labeled by ($n,m$) and wave functions on the two ends are denopted by $\psi_{n,m}$(black circles) and $\phi_{n,m}$(white circles).
The hopping integrals for the horizontal bonds are  $t$ and those for the others are $1$. }
\label{HoneyLattice}
\end{center}
\end{figure}

The honeycomb lattice consists of two sublattices A and B which are the triangular lattice. A point in sublattice A and a point in sublattice B are not equivalent (see Fig. \ref{HoneyLattice}). Thus it is non-Bravais and the dual space has a quite non-trivial structure. 

In this paper we study the tight-binding model on the honeycomb lattice. The electronic states are Bloch waves which have a peculiar energy spectrum with the $2d$ Dirac zero modes and also the $1d$ Dirac modes which have not been well-recognized. A $U(1)$ gauge field is constructed from the Bloch waves. This kind of gauge field was originally introduced in $2d$ periodic systems in a magnetic field in the study of quantum Hall effect(QHE)\cite{ann}. When time-reversal symmetry is broken, the gauge field is topologically non-trivial in the magnetic dual space and Chern numbers of the pertinent fiber bundle give quantized Hall conductance.

In Sec. \ref{TBM},  the tight-binding model is introduced which represents nearest-neighbor electron hopping on the honeycomb lattice. The Bloch states and the energy spectrum are given. In Sec. \ref{DualSpace}, the dual space for the Bloch states are discussed. A convention to use two Brillouin zones is taken, and it is shown that it has a non-trivial topology being a torus $T^2$. In Sec. \ref{DiracZeroMode}, the Dirac zero modes at $E=0$ are discussed. Especially it is shown that Bloch's theorem used in Sec. \ref{TBM} is no longer valid and the states are high degenerate. These Dirc modes are two-dimensional, but in Sec. \ref{1dDirac} the existence of the $1d$ Dirac modes is shown. The corresponding states are again highly degenerate and it is possible to  construct the strap dimer states which are strictly localized in one direction and extended in the other direction.

In Sec. \ref{GaugeField}, the $U(1)$ gauge field for Bloch electrons is introduced  and then, in Sec. \ref{super}, superconductivity with quantized Abrikosov fluxes in the dual space is discussed.  In Sec. \ref{MassiveCase}, a massive case is considered in which there is no Dirac zero modes. Robustness of the fluxes is shown.  In Sec. \ref{FluxMonopole}, the $2d$ dual space is extended to $3d$. In the $U(1)$ gauge theory there is a  magnetic flux line  but there is no magnetic monopoles. On the other hand, there exists a monopole-antimonopole pair  in the non-Abelian $SO(3)$ theory which shows confinement.  In this case, the $SO(3)$ gauge group  is broken down to $U(1)$ at the 2$d$ boundary.
In Sec. \ref{HallConductance},  the topological expression in QHE relates this monopole confinement and time reversal symmetry. In Sec. \ref{QHEDeconfinement}, QHE of the honeycomb lattice is discussed. Energy spectra against magnetic fluxes per unit cell of the honeycomb lattice are shown for a number of anisotropic cases. They are compared with that of the square lattice.  The extended $3d$ dual space is split into $q$ subsapces and it is shown that monopoles are deconfined due to breaking of time reversal symmetry.

In the final Section \ref{JahnTeller}, the Jahn-Teller effect is briefly discussed.

\section{Tight-binding model on the honeycomb lattice}
\label{TBM}
The honeycomb lattice is an alternate lattice  which consits of sublattice A and  sublattice B. The two sublattices are triangular lattices (see Fig. \ref{HoneyLattice}).  Since a point in sublattice and a point in sublattice B are not equivalent the honeycomb lattice is not a Bravais lattice. Most of  the nontrivial results which are discussed in this paper are immediate consequences of this fact.

Denote wave functions on sublattices A and B as  
 $\psi_{n,m}$ and $\phi_{n, m}$, respectively, as shown in Fig. \ref{HoneyLattice}.
The tight-binding model with  nearest neighbor hopping is
\be
- \phi_{n+1, m-1} - \phi_{n+1, m+1 }- t\phi_{n,m}&= & E \psi_{n,m},
\nonumber \\
- \psi_{n-1, m+1} - \psi_{n-1, m-1} - t\psi_{n,m} &=&E \phi_{n,m},
\label{tb1}
\ee
where hopping integrals of the horizontal bonds are $t$ and those for the other bonds are $1$.

Sublattice A and sublattice B are the triangular Bravais lattices and each has a set of primitive vectors
\be
{\bm a_1} &=& (1, -1), \nonumber \\
{\bm a_2} &=& (1, 1).
\label{dvector}
\ee
Then the  reciprocal lattice can be generated by
\be
{\bm b_1} &=& \pi (1, -1), \nonumber \\
{\bm b_2} &=& \pi (1, 1),
\label{rvector}
\ee
where 
\be
{\bm b_i} \cdot {\bm a_j} = 2\pi \delta_{ij}.
\ee

Since the system is periodic,  Bloch's theorem applies and the wave functions are written 
\be 
\psi_{n,m}& =& e^{ik_x n+ik_ym} u({\bm k}), \nonumber \\
\phi_{n, m} & = & e^{ik_x n+ik_y m} v({\bm k}).
\label{bloch}
\ee
Here the functions $u({\bm k})$ and $v({\bm k})$ are periodic in the dual space, namely
\be
u({\bm k} + {\bm R}) = u({\bm k}), \nonumber \\
v({\bm k} + {\bm R}) = v({\bm k}),
\ee
if ${\bm R}$ is a reciprocal lattice vector which is  given by
\be 
{\bm R} = k_1{\bm b_1} + k_2 {\bm b_2},
\ee
with integers $k_1$ and $k_2$.

Equations (\ref{tb1}) and (\ref{bloch}) give
\be
h({\bm k})\left[ \begin{array}{c} u({\bm k}) \\ v({\bm k})
\end{array} \right] = E\left[ \begin{array}{c} u({\bm k}) \\ v({\bm k})
\end{array} \right],
\label{tb-2}
\ee    
where the $2 \times 2$ matrix hamiltonian is 
\be
h({\bm k})= \left[ \begin{array}{cc} 0 & \Delta ({\bm k}) \\ \Delta^*({\bm k}) &0
\end{array} \right],
\label{hamil}
\ee
with
\be
\Delta({\bm k}) &=& -t-2e^{ik_x} \cos k_y.
\label{delta}
\ee
From Eqs.(\ref{tb-2}) and (\ref{hamil}) we have
\be
 \frac{u({\bm k})}{ v({\bm k})} = \pm \frac{\Delta({\bm k})}{|\Delta({\bm k})|},
\label{wf}
 \ee
and
\be
E = \mp |\Delta({\bm k})|.
\label{energy}
\ee
The energy spectrum for the isotropic case, $t = 1$, is shown in Fig. 2 and it is symmetric with respect to $E=0$. This is the consequence of the fact that the tight-binding model (\ref{tb1}) is bipartite.
\begin{figure}[t]
\begin{center}
\includegraphics[width=55mm]{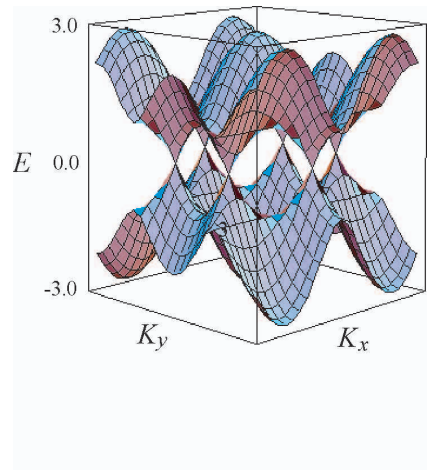}
\caption{(Color online) The energy spectrum of the honeycomb lattice.}
\end{center}
\label{HoneyE}
\end{figure}

\section{Topological structure of the dual space}
\label{DualSpace}
\begin{figure}[t]
\includegraphics[width=35mm]{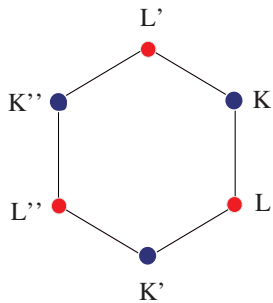}
\caption{(Color online) The first Brillouin zone of the honeycomb lattice and the Dirac zero modes. Dirac points K, K', and K'' are equivalent since they differ by reciprocal lattice vectors each other. In the same manner Dirac points L, L', and L'' are equivalent.}
\label{fbz}
\end{figure}
A metric of the lattice is  not  included in the last section.  Thus $k_x$ and $k_y$ are dimensionless. Let us introduce a length scale $a$ and scale $k_x$ and $k_y$ by,
\be
K_x &=& \frac{2}{\sqrt{3}a} k_x, \nonumber \\
K_y &=& \frac{2}{a} k_y.
\label{rescale}
\ee
In this scale the direct lattice has the hexagonal symmetry and the honeycomb lattice consists of two triangular sublattices whose lattice spacings are $a$. The primitive vectors of the reciprocal lattice (\ref{rvector}) are scaled and given by
\be
{\bm b_1}' &=& (\frac{2\pi}{\sqrt{3} a}, -\frac{2\pi}{a}), \nonumber \\
{\bm b_2}' &=& (\frac{2\pi}{\sqrt{3} a}, \frac{2\pi}{a}).
\label{rvector2}
\ee
The reciprocal lattice is generated  as 
\be 
{\bm R}' = k_1{\bm b_1}' + k_2 {\bm b_2}'
\ee
where $k_1$ and $k_2$ are integers.
The first Brillouin zone is the Wigner-Seitz cell of the reciprocal lattice and it is shown in Fig. \ref{fbz}.

A Brillouin zone is defined only for a Bravais lattice. So note that the first Brillouin zone here, in fact, is for sublattice A (or B) which is the triangular lattice.

The corners of the first Brillouin zone are given by
\be
{\rm K} : & ( \frac{2\pi}{\sqrt{3} a},  \frac{2\pi}{3a}), & {\rm L}:  ( \frac{2\pi}{\sqrt{3} a}, -\frac{2\pi}{3a}), \nonumber \\
{\rm K'} : & (  0,  - \frac{4\pi}{3a}),& {\rm L'} :  (  0,  \frac{4\pi}{3a}), \nonumber \\
{\rm K''} :  & (-\frac{2\pi}{\sqrt{3} a},  \frac{2\pi}{3a}), & {\rm and } \; {\rm L''} :   (-\frac{2\pi}{\sqrt{3} a}, - \frac{2\pi}{3a}).
\label{KL}
\ee

\begin{figure}[t]
\includegraphics[width=80mm]{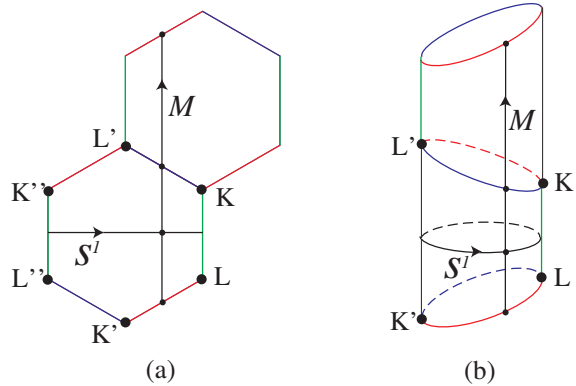}
\caption{(Color online) (a) The dual space which consists of two Brillouin zones. Green lines, red lines, and blue lines are equivalent, respectively for themselves. The line  in $x$-direction is topologically $S^1$, and the line  in $y$-direction $M$ is also $S^1$.   (b) The dual space with non-trivial  topology. }
\label{MS}
\end{figure}

In Fig. \ref{MS} (a), two Brillouin zones are drawn. 
These two are equivalent, but this redundant representation is convenient. This can be seen as follows: The vertical zone boundaries (green lines) of the first Brillouin zone are equivalent since they differ by a reciprocal lattice vector $(4\pi/\sqrt{3}a, 0)$. 
Therefore  a horizontal line is a circle $S^1$. On the other hand the zone boundaries of the first Brillouin zone connected by a vertical line are not equivalent.  The vertical line, $M$, reaches  an equivalent point only after going through the two Brilloin zones. 

The dual space is a torus, $S^1\times M$, which is shown in Fig. \ref{MS} (b). The top and the botom surfaces of $S^1 \times M$ have to be identified since they are related by a reciprocal lattice vector $(0, 4\pi/a)$.

A direction in the dual space is equivalent to ones which are rotated by $\pm 2\pi/3$ due to the hexagonal symmetry of the honeycomb lattice. Thus we call $x$-direction and the ones which are rotated by $\pm 2\pi/3$ as the $S^1$ directions. In a similar manner, $y$-direction and the ones which are rotated by $\pm 2\pi/3$ are called $M$ directions.

\section{Dirac zero modes}
\label{DiracZeroMode}
At a glance a zero energy state seems to be given by $\Delta({\bm k}) =0$ from Eq. (\ref{energy}). Note that, if so, $h({\bm k})$ in Eq. (\ref{hamil}) vanishes and the problem is ill-defined. 

Facing this situation let us go back to Eq. (\ref{tb1}) before applying Bloch's theorem (\ref{bloch}). For $E=0$, $\psi_{n,m}$ and $\phi_{n,m}$ are decoupled and Bloch's theorem (\ref{bloch}) is no longer valid. Namely there is no phase relation between $\psi_{n,m}$ and $\phi_{n,m}$ because they are solutions of two independent equations,
\be
\psi_{n-1, m+1} + \psi_{n-1, m-1} + t\psi_{n,m} &=& 0, \nonumber \\
\phi_{n+1, m-1} + \phi_{n+1, m+1 } + t\phi_{n,m} &=& 0.
\label{zeromode}
\ee
Bloch's theorem is valid for each equation separately and Eq.(\ref{zeromode}) is satisfied if
\be
k_x &= 0,\, \cos k_y &= -\frac{t}{2}, 
\label{dirac1}
\ee
or
\be
k_x &= \pm \pi,\, \cos k_y& = \frac{t}{2}.
\label{dirac2}
\ee

Note that there is no Dirac point if 
\be
t > t^* = 2.
\label{tc}
\ee

For the isotropic case, $t=1$, we have 
\be
k_x &= 0,\, k_y &= \pm \frac{2\pi}{3},
\label{dirac3}
\ee
or
\be
k_x &= \pm \pi,\, k_y& = \pm \frac{\pi}{3}.
\label{dirac4}
\ee
These points correspond to the corner of the first Brillouin zone (\ref{KL}) (see Fig. \ref{fbz}).

Two degenerate states are
\be
\psi_{n,m}({\bm k}) &=& e^{k_x n + k_y m}, \nonumber \\
 \phi_{n,m} ({\bm k}) &=& 0,
\ee
and
\be
\psi_{n,m}({\bm k}) &=& 0, \nonumber \\
 \phi_{n,m} ({\bm k}) &=& e^{k_x n + k_y m}.
\ee
The first state is nonzero only on sublattice A and the second  state is nonzero only on sublattice B. Since they are degenerate, any linear combination of the two states is an eigenstate, namely
\be
f_{n,m}({\bm k}) = \alpha({\bm k}) \psi_{n,m}({\bm k}) + \beta({\bm k})\phi_{n,m} ({\bm k})
\label{wfzm}
\ee
 is an eingenstate for arbitarary functions  $\alpha({\bm k})$ and   $\beta({\bm k})$. Thus the states are highly degenerate.

The key point is that non-Bloch forms may be taken for the eigenstates at the Dirac zero modes.

\section{One-dimensional  Dirac modes and the strap dimer states}
\label{1dDirac}
\begin{figure}[t]
\begin{center}
\includegraphics[width=30mm]{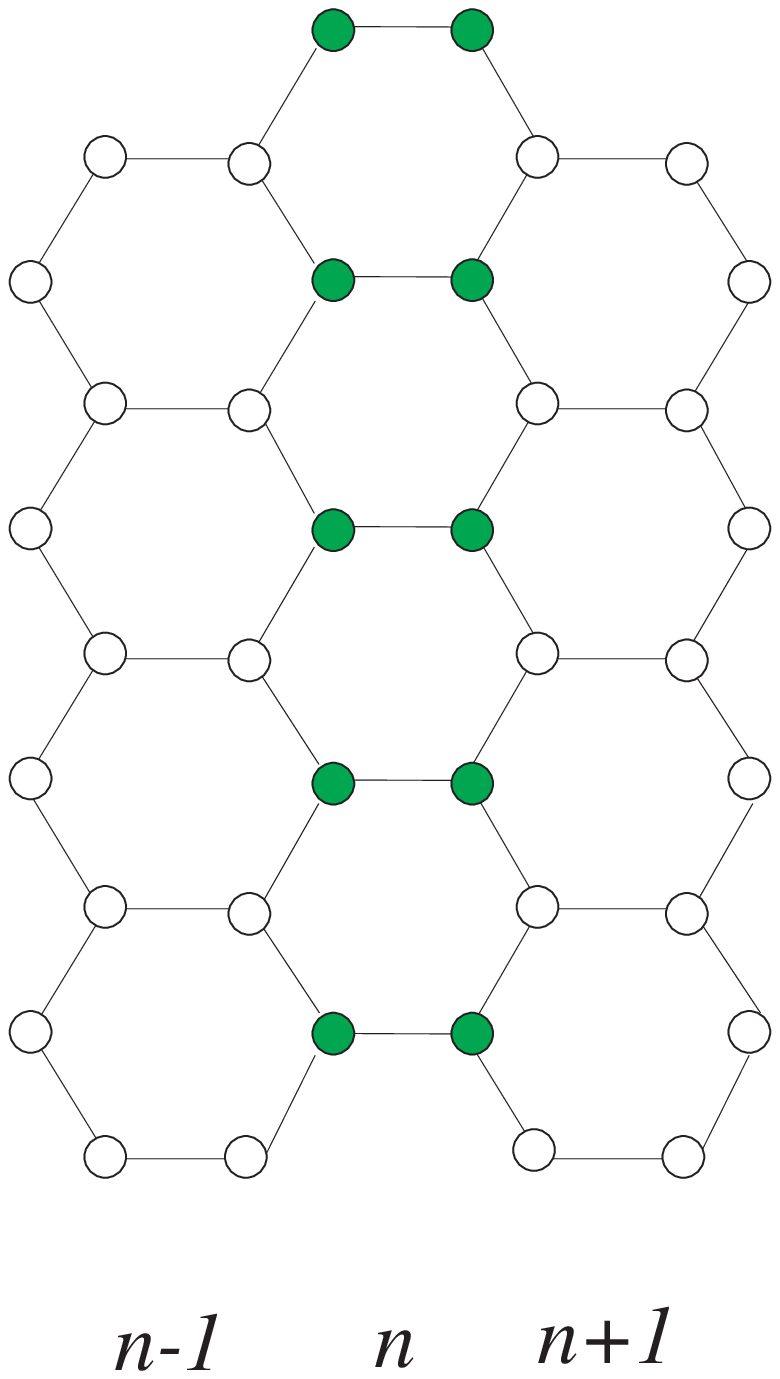}
\caption{(Color online) A strap dimer state. It is only nonzero on the green circles. }
\label{StrapDimer}
\end{center}
\end{figure}

If $k_y = \pm \pi/2$, Eq. (\ref{delta}) gives $\Delta =-t$ which is independent of $k_x$. Instead of Eq.(\ref{bloch}), let us write
\be
\psi_{n,m} = e^{ik_y m} \psi_n \nonumber \\
\phi_{n,m} = e^{ik_y m} \phi_n,
\label{1dBloch}
\ee
then Eq. (\ref{tb1}) reduces to
\be
 -t \left[ \begin{array}{cc} 0 & 1 \\ 1 &0\end{array} \right]  \left[ \begin{array}{c} \psi_n \\ \phi_n
\end{array} \right] =E \left[ \begin{array}{c} \psi_n \\ \phi_n
\end{array} \right].
\label{dimer}
\ee
Here $n$ is arbitrary, and we have 
\be
 \phi_n &=& \pm \psi_n, \nonumber \\
 E &=&\mp t.
\ee 
A state ($ \phi_n,  \psi_n$) is shown in Fig. \ref{StrapDimer} and it is  strictly localized in $x$-direction and extended in $y$-direction. 

States with different $n$'s are degenerate, and therefore any linear combinations of them are also eigenstates. In particular, we choose $\psi_n = e^{ik_x n}, \phi_n = \pm e^{ik_x n}$ in Eq. (\ref{1dBloch})   and
\be
\psi_{n,m} =e^{ik_xn} e^{ik_y m},  \nonumber \\
\phi_{n,m} = \pm e^{ik_xn} e^{ik_y m}.
\ee
These staes are in a Bloch's form and extended in both $x$- and $y$-directions.
From Eqs.(\ref{delta}) and (\ref{energy}), Fermi velocities of these 1$d$ Dirac modes are given by
\be
v_F = \pm 2 \cos k_x.
\label{1dvf}
\ee

\section{Gauge fields for Bloch electrons}
\label{GaugeField}
Gauge fields for Bloch electrons were constructed  first  in the context of quantum Hall effect where time reversal symmetry is broken by an external magnetic field \cite{ann}. Although time reversal symmetry is not broken here let us  define, nonetheless, a field  in the dual space by
\be
A({\bm k})=-i [ u({\bm k})^* \nabla_k u({\bm k}) + v({\bm k})^* \nabla_k v({\bm k}) ].
\label{gauge}
\ee
In order to clarify the gauge structure, consider overall phases of wave functions. If $(u({\bm k}), v({\bm k})) $ is an eigenvector of $h({\bm k})$, so is 
\be
(u({\bm k})', v({\bm k})' ) =e^{if({\bm k})} (u({\bm k}), v({\bm k})).
\ee
 The field (\ref{gauge}) is transformed as
\be
A({\bm k})' = A({\bm k})+ \nabla_k  f({\bm k}).
\label{gauge-tr}
\ee
Thus there is a  formal equivalence between $A({\bm k})$ and the vector potential in electromagnetism. Therefore we call $A({\bm k})$ as a ``gauge field" in the dual space. The gauge invariance comes from the freedom of choosing overall phases of wave functions.  A ``magnetic field" in the dual space is given by 
\be
B({\bm k}) = \nabla_k \times A({\bm k}),
\label{bdual}
\ee
which is gauge-invariant. 

At the Dirac points we have even stronger gauge invariance. For a wave function in a form of (\ref{wfzm}) one may change phases of $\alpha({\bm k})$ and   $\beta({\bm k})$ separately. Thus we have independent gauge symmetries on sublattice A and sublattice B. This phenomenon is caused by {\em breaking of Bloch's theorem}.

\section{Superconductivity and quantized Abrikosov fluxes}
\label{super}
Take ratio of  wave functions  on sublattice A and sublattice B,
\be
z({\bm k}) = \frac{\psi_{nm}}{\phi_{nm}} .
\label{z1}
\ee
We have $z({\bm k}) = u({\bm k})/ v({\bm k})$ from the Bloch form of wave functions (\ref{bloch}) and Eq. (\ref{wf}) gives
\be 
z({\bm k})=\pm \frac{\Delta({\bm k})}{|\Delta({\bm k}) |}.
\label{z2}
\ee
Thus write
\be
u({\bm k}) &=&  z({\bm k}) e^{ig({\bm k})}, \nonumber \\
 v({\bm k}) &=&  e^{ig({\bm k})},
 \label{uv}
 \ee
and 
\be
z({\bm k}) = e^{i\theta({\bm k})}.
\label{z3}
\ee
From Eq. (\ref{uv}), the gauge  field (\ref{gauge}) is given by 
$A({\bm k})= -i z^*({\bm k}) \nabla_k z({\bm k}) +  2 \nabla_k g({\bm k}) $.
Since the second term is a pure gauge we can write, in a suitable gauge,
\be
A({\bm k})=-i z^*({\bm k}) \nabla_k z({\bm k}) = \nabla_k \theta({\bm k}),
\label{a-flux}
\ee
where Eq. (\ref{z3}) is used.
This shows that gauge field $A({\bm k})$ is an exact $1$-form in tems of  $\theta({\bm k})$ which are phase differences between wave functions on sublattiuce A and sublattice B.

In this manner we have a $U(1)$ gauge field theory and then  it is natural to interpret it as a theory of superconductivity.
Let us identify  $\Delta({\bm k})$  as the Ginzburg-Landau order parameter for  superconductivity in the dual space. Then London current  is $j_\nu({\bm k}) = -m_s^2 A_\nu({\bm k})$, where $m_v$ is mass of the $U(1)$ field. For the magnetic field in the dual space (\ref{bdual}) we have
\be
\int_S B({\bm k})dS &=& \int_S \nabla \times A({\bm k}) dS \nonumber \\
&=&  \oint_C A({\bm k})d\ell = \oint_C \nabla_k \theta({\bm k}) = 0,
\ee
unless  a coutour $C$ contains a Dirac point. This is {\em Meissner effect}; a magnetic field can not  penetrate into bulk of a superconductor.

If a contour C includes  a Dirac point we have
\be
\int_S B({\bm k})dS &=& 2\pi, \; {\rm if} \; C \ni {\rm K}, \nonumber \\
\int_S B({\bm k})dS &=& - 2\pi, \;{\rm if} \; C \ni {\rm L},
\label{flux}
\ee
for both the upper and the lower bands. Thus we have  flux $2\pi$ at K and flux $-2\pi$ at L\cite{FluxUnit}.
These are quantized Abrikosov fluxes. 
The states are in {\em Type II superconducting  with Abrikosov fluxes at the Dirac points}. The total Abrikosov flux is zero for either the  upper or the lower band.

\section{Abrikosov Fluxes in a Massive Case}
\label{MassiveCase}
Let us include potentials $m$ on sublattice A and $-m$ on sublattice B. 
 Hamiltonian (\ref{hamil}) is modified as 
\be
h({\bm k})= \left[ \begin{array}{cc} m & \Delta ({\bm k}) \\ \Delta^*({\bm k}) & -m
\end{array} \right].
\label{hamil2}
\ee
The eigenstates and energies are given by
\be
 \frac{u({\bm k})}{ v({\bm k})} = z({\bm k}) =  \frac{\Delta({\bm k})}{E -m},
\label{wf2}
 \ee
and
\be
E = \pm \sqrt{ m^2 + |\Delta({\bm k})|^2}.
\label{energy2}
\ee
Since phases of  $z({\bm k})$ remain the same, it can be shown that Eq. (\ref{flux}) is still valid. 
Thus we have quantized Abrikosov fluxes at the same points K and L with fluxes $2\pi$ and $-2\pi$, respectively.  

Note, however, that there are no Dirac points at the Abrikosov fluxes. Therefore we conclude that the existence of Abrikosov fluxes is nothing to do with Dirac zero modes.

\section{Magnetic Fluxes and Monopoles}
\label{FluxMonopole}
\begin{figure}[t]
\includegraphics[width=55mm]{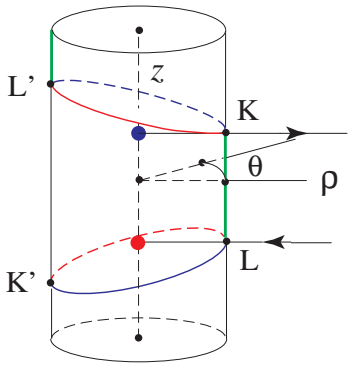}
\caption{(Color online) The extended 3$d$  dual space $R^3 = (\rho, \theta, z)$. A $2\pi$ Abrikosov flux is at K and a $-2\pi$ Abrikosov flux is at L. Points K' and L' are equivalent to K and L, respectively. A monopole is at the blue circle and an antimonopole is at the red circle  in the non-Abelian $SO(3)$ gauge theory. The 3$d$  dual space $R^3$ is periodic in $z$-direction with period $4\pi/a$.}
\label{R3}
\end{figure}

 In Sec. \ref{DualSpace}, it is shown that the dual space is a torus $T^2$. We extend it to a 3$d$ space $R^3$ which is shown in Fig. \ref{R3}.
  This 3$d$ space is periodic in $z$-direction with period $4 \pi/ a$. The 2$d$ dual space is the surface of the cylinder with radius $\rho = 2/\sqrt{3}a$. 

Let us define the Ginzburg-Landau Lagrangian  in $R^3$:
\be
L = \int d^3 k [\frac{1}{4}  F_{ij}^2 + \frac{1}{2} |D_i {\bf \varphi}|^2 + a |\varphi|^2 + \frac{b}{2} |\varphi|^4 + \cdots ],
\label{LU1}
\ee
where
\be
F_{ij}& = & \epsilon^{ij}\partial_i A_j,  \nonumber \\
\bf {\varphi} &=& (\varphi^1,\: \varphi^2),  \nonumber \\
D_i \varphi^a &=& (\partial_i -i   A_i) \varphi^a .
\ee
The Lagrangian (\ref{LU1}) is invariant under a $U(1)$ gauge transformation:
\be
\varphi \rightarrow e^{i\Lambda} \varphi, \nonumber \\
A_i \rightarrow A_i + \partial_i \Lambda.
\ee
The potential term is chosen to give the field configuration in the 2$d$ dual space which is  described in Sec. \ref{super}.

In the 2$d$ dual space there is a  $2\pi$ Abrikosov flux at K and a $-2\pi$ Abrikosov flux at L'. In the extended space, $R^3$, a flux line cannot have an end point due to topological stability  of the Abelian $U(1)$ gauge theory. Also a flux line costs an energy which is propotional to its length. From these it can be seen that  a magnetic flux line  is a straight line which goes through from L' to K. (There is another flux line which goes through from L to K', but this is equivalent to the former.) Thus there is no magnetic monopole.

A much more interesting subject is the non-Abelian gauge theory.
Let us consider the $SO(3)$ gauge symmetry which has $U(1)$ as a subgroup.
\be
L = \int d^3 k [\frac{1}{4} ( F_{ij} )^2 + \frac{1}{2} ( D_i {\bf \varphi} )^2 + V({\bf \varphi}) ],
\ee
where
\be
F_{ij}^a & = & \partial_i A_j^a - \partial_j A_i ^b - \epsilon^{abc}A_i^b A_j^c,  \nonumber \\
{\bf \varphi} &=& (\varphi^1,\: \varphi^2,\: \varphi^3), \nonumber \\
D_i \varphi^a &=& \partial_i \varphi^a - \epsilon^{abc} A_i^b \varphi^c.
\ee
The potential $V({\bf \varphi})$ is chosen to give the $U(1)$ field configuration on the 2$d$ dual space which is described in Sec. \ref{super}. Thus the $SO(3)$ gauge symmetry is broken down to the subgroup $U(1)$ on the surface. 

This symmetry breaking is forced on the surface and is {\em not} spontaneous as in the spontaneously broken gauge theory of  't Hooft\cite{thooft} and Polyakov. The $U(1)$ gauge structure of Bloch electrons on the honeycomb lattice determine the symmetry on the  2$d$ surface and they break the $SO(3)$ gauge symmetry.

As shown by  't Hooft\cite{thooft} monopoles are allowed in the $SO(3)$ gauge theory. From the symmetry of the dual space $R^3$ we conclude that there exists a pair of monopole and antimonopole as shown in Fig. \ref{R3}.

\section{Hall Conductance and  Monopole Confinement}
\label{HallConductance}
The Hall conductance in units of $e^2/h$ from a single band is given by the topological expression:
\be
\sigma_{xy}= \frac{1}{2\pi} \int d^2 k [\nabla_k \times A({\bm k})]_z,
\label{hall}
\ee
where the integral is over the first Brillouin zone\cite{ann}. This is always an integer and called a TKNN integer\cite{tknn}. In the present problem  we do not expect that the Hall effect takes place because there is no magnetic field and time reversal symmetry is not broken.

On the other hand, as shown above, there exists a pair of monople and antimonople. There is neither an isolated monopole nor an isolated antimonopole. This fact and the vanishing of the Hall conductance is related by the topological expression of Hall conductance (\ref{hall}). The R.H.S represents the total flux (sum of the fluxes at Dirac points K and L) through the $2d$ dual space.

Therefore we conclude that {\em the monopole and antimonopole pair confinement is forced by time reversal symmetry}.

\section{Quantum Hall effect and monopole deconfinement}
\label{QHEDeconfinement}
\begin{figure}[t]
\begin{center}
\begin{tabular}{c}
\includegraphics[width=55mm]{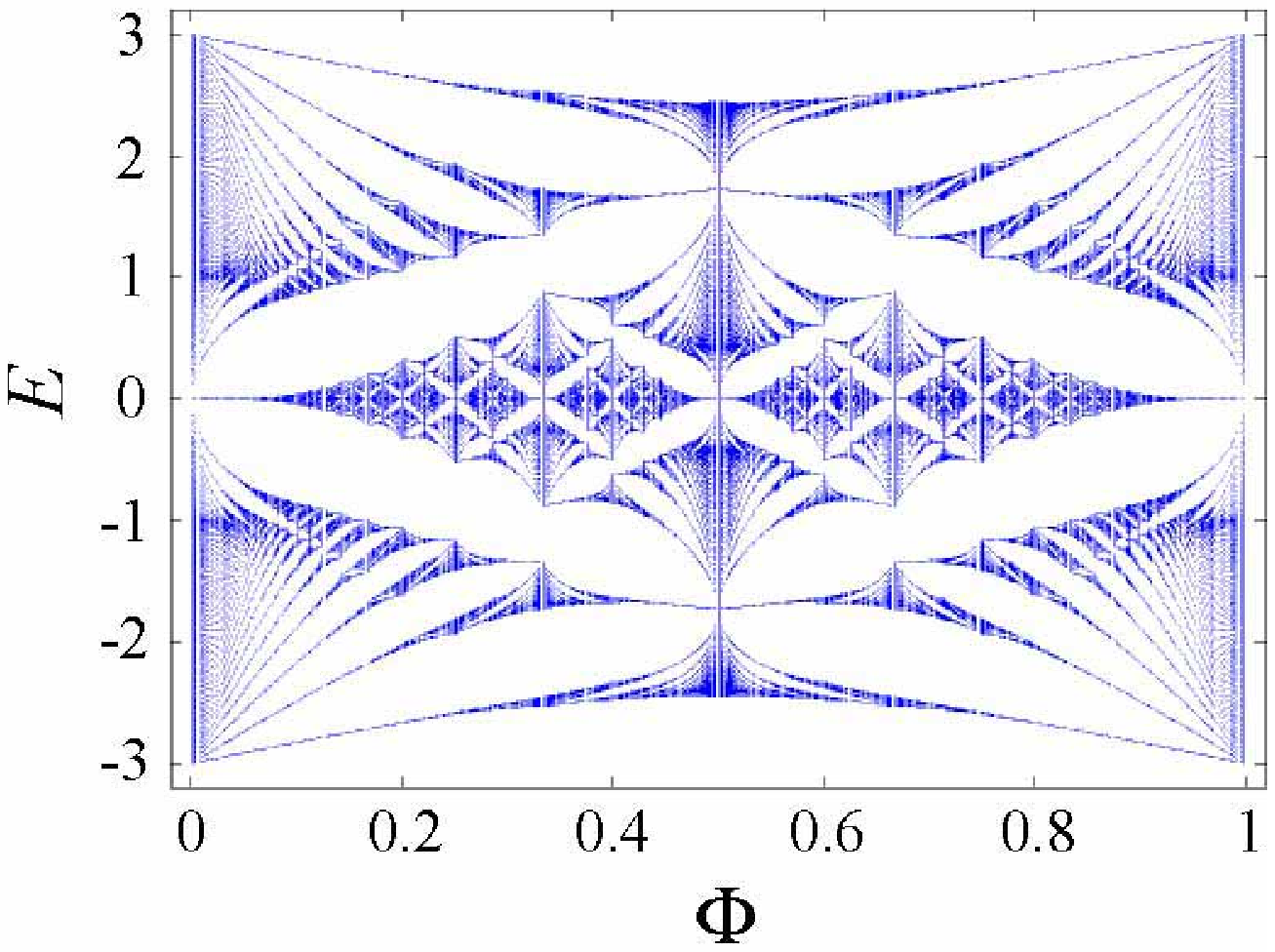}
(a)
\\
\\
\includegraphics[width=55mm]{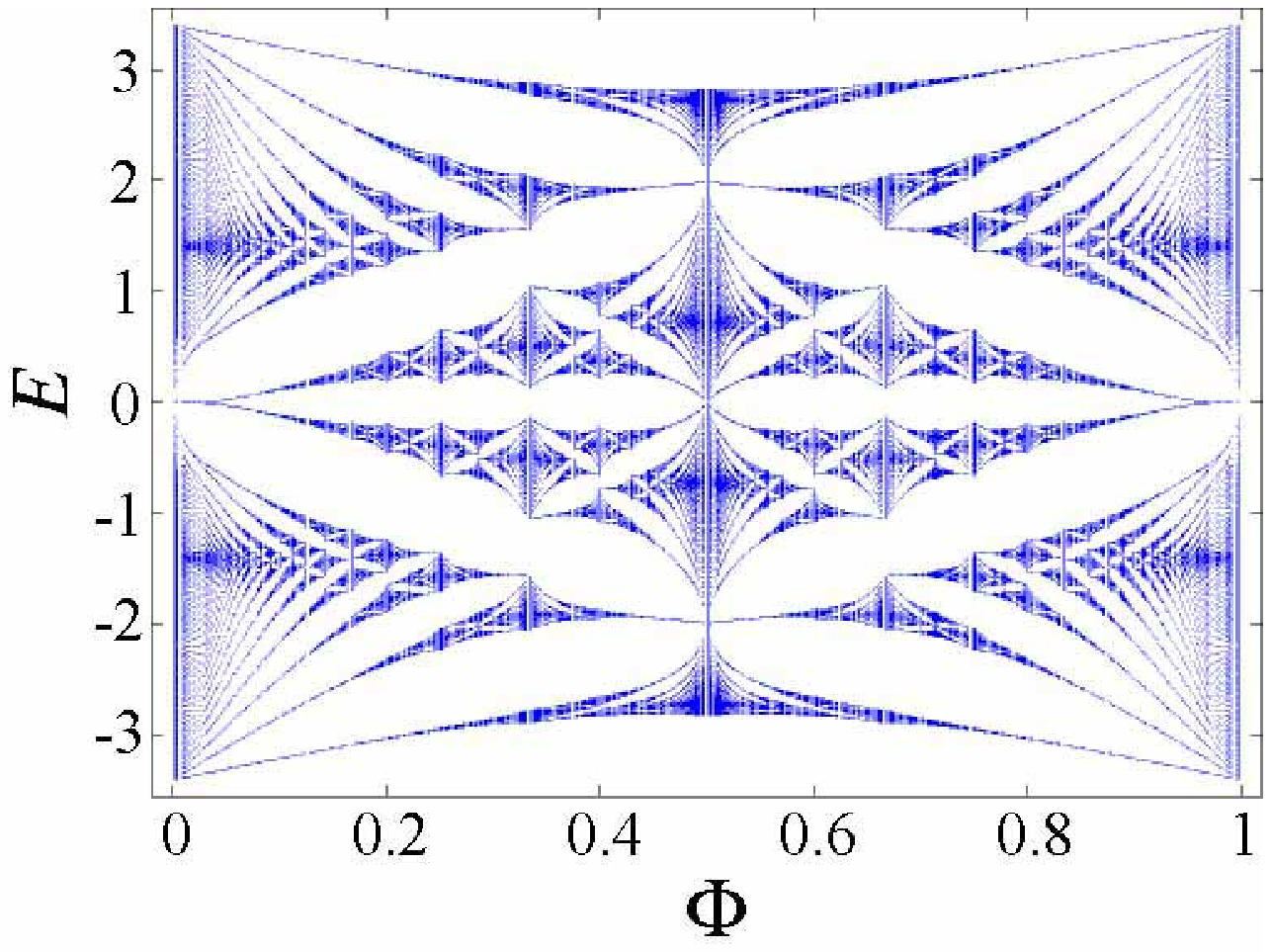}
(b)
\\
\\
\includegraphics[width=55mm]{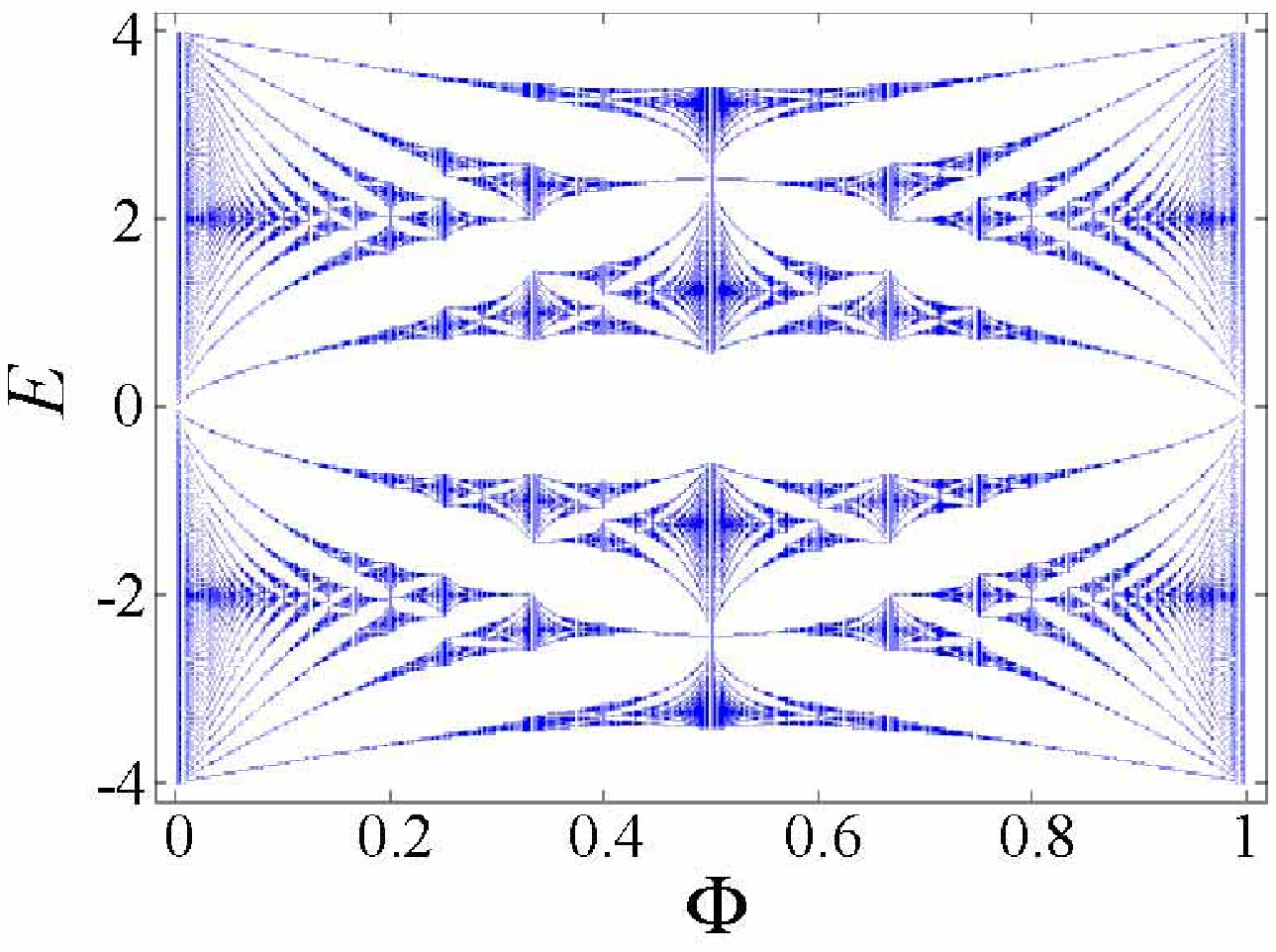}
(c)
\\
\\
\includegraphics[width=55mm]{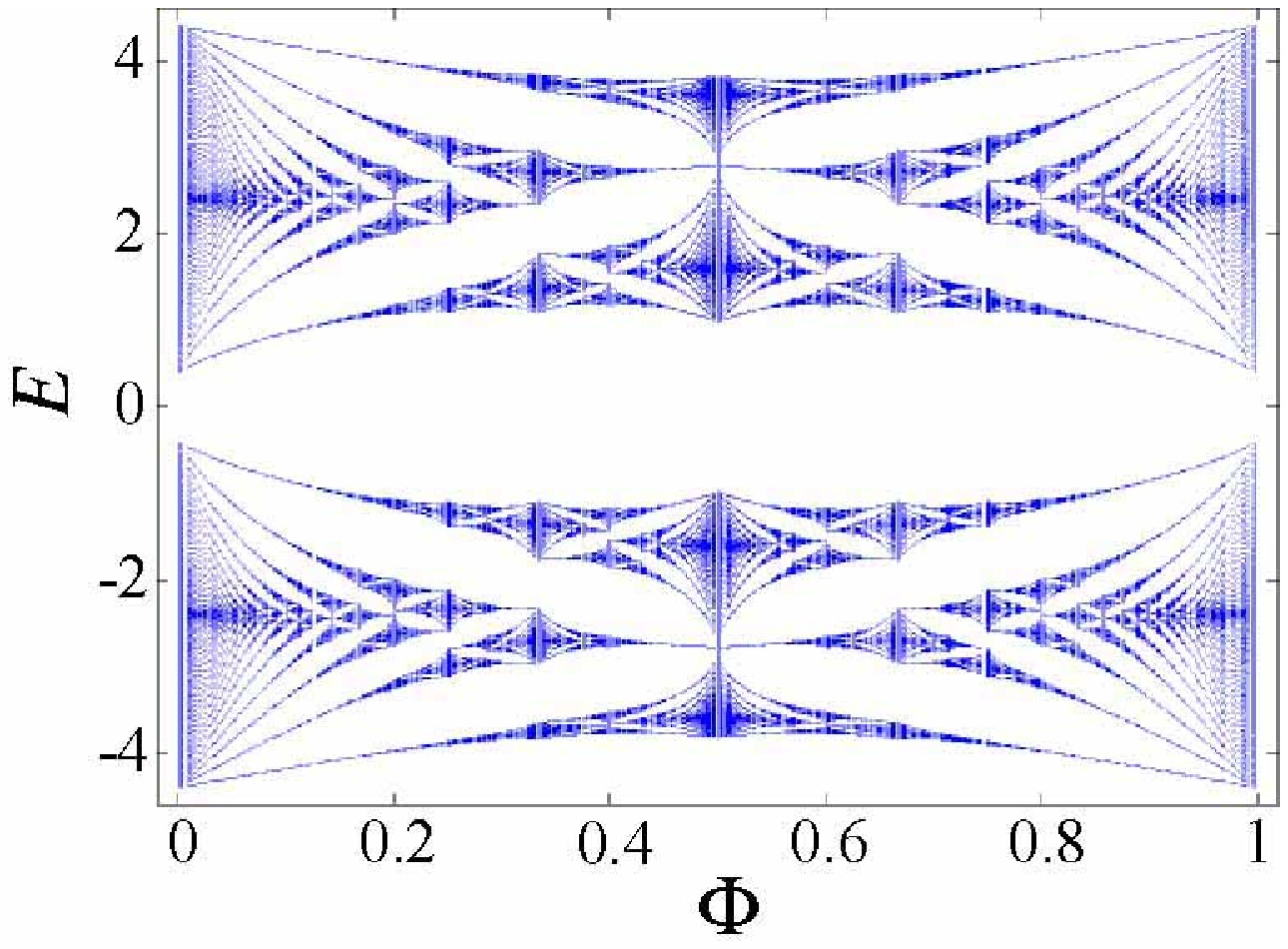}
(d)
\end{tabular}
\caption {(Color online) The energy spectra of the honeycomb lattice in magnetic fields for (a) $t = 1.0$, (b) $t = 1.4$, 
(c) $t = 2.0$, and (d) $t = 2.4$. A magnetic flux through a unit cell is given by $2\pi \Phi$. }
\label{HoneyHofst}
\end{center}
\end{figure}

\begin{figure}[t]
\includegraphics[width=55mm]{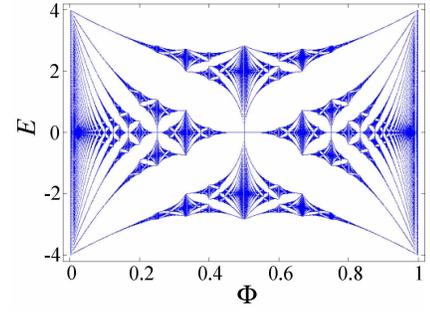}
\caption{(Color online) The energy spectrum of the square lattice in a magnetic field.}
\label{sq}
\end{figure}

In order to describe electron hopping on the honeycomb lattice in a magnetic field, phase factors are introduced in  Eq. (\ref{tb1}),
\be
- \phi_{n+1, m-1} - \phi_{n+1, m+1 }- te^{-i \pi \Phi m}\phi_{n,m}=  E \psi_{n,m},
\nonumber \\
- \psi_{n-1, m+1} - \psi_{n-1, m-1} - t e^{i \pi \Phi m}\psi_{n,m} =E \phi_{n,m}.
\label{tbmag}
\ee
Here a magnetic flux through a unit hexagon is given by $2\pi \Phi$. The energy spectra are shown in Fig. \ref{HoneyHofst} for (a) $t = 1.0$, (b) $t = 1.4$, (c) $t = 2.0$, and (d) $t = 2.4$.
If there is no magnetic field, $\Phi = 0$, the spectra consist of an upper band and a lower band. As shown in Sec. \ref{DiracZeroMode}, the two bands touch at the Dirac points for $t < t_c = 2$. This situation is similar to the case of the square lattice with $\Phi = 1/2$ (see  Fig. \ref{sq}). For $t > t_c = 2$ there is no Dirac point but there is a gap around $E = 0$ (see Fig. \ref{HoneyHofst} (d)).

If $\Phi = p/q$, where $p$ and $q$ are mutually prime integers, the translational symmetry of the honeycomb lattice in $y$-direction is broken. Equation (\ref{tbmag}) becomes periodic with period $q$ in $y$-direction. As a consequence the 2$d$ dual space is spilt into $q$ subspaces and both the upper and the lower bands are split into  $q$ subbands (see Fig. \ref{HoneyHofst}). This phenomena has been known for the square lattice (Fig. \ref{sq}) and let us call it as the {\em Hofstadter mechanism} of band splittings\cite{hofst}.

When the Fermi energy is in a gap, the Hall conductance is given by a sum of contributions from the subbands below the Fermi energy. A contribution from a single subband is given by the topological expression (\ref{hall}) and thus quantized. Quantized Hall effect of the honeycomb lattice following these topological considerations is discussed in Ref. \cite{HasegawaKohmoto}.

\begin{figure}[t]
\begin{center}
\includegraphics[width=55mm]{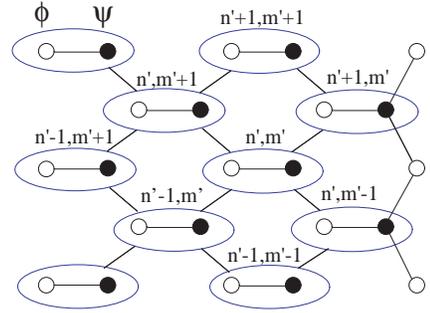}
\caption{(Color online) The square lattice for  bonding states and ant-bonding states.}
\label{SquareLattice}
\end{center}
\end{figure}

If $t > t_c = 2$, there is a gap around  $E = 0$  with zero Hall conductance (see Fig. \ref{HoneyHofst}(d)). In fact, both the upper and the lower parts of the spectrum consist of $q$ subbands and they  are topologically equivalent, respectively, to that of the square lattice which is shown in Fig. \ref{sq}. 
In order to see this, first consider the case $\Phi = 0$.
Take sum and difference of the two equations in Eq. (\ref{tb1}),
\be
&\phi_{n+1, m-1} + \phi_{n+1, m+1 } +\psi_{n-1, m+1} + \psi_{n-1, m-1} \nonumber \\
&=  -(E +t) (\psi_{n,m} + \phi_{n,m}),
\nonumber \\
&\phi_{n+1, m-1} + \phi_{n+1, m+1 } - \psi_{n-1, m+1} - \psi_{n-1, m-1} \nonumber \\
&=  -(E-t) (\psi_{n,m}-\phi_{n,m}).
\label{tb2}
\ee
For $t \gg 1$ we have bonding states, $\phi_{n, m } =\psi_{n, m}$, and anti-bonding states, $\phi_{n, m } = -\psi_{n, m}$. For those states Eq. (\ref{tb2}) represents a tight-binding model on the square lattice which is 
 shown in Fig. \ref{SquareLattice}.
The solutions for those two sectors are: for bonding states,
\be
\psi_{n',m'} &=& \phi_{n',m'}=e^{ik_1n' + ik_2m'}, \nonumber \\
E &=& -t - (\cos k_1 + \cos k_2),
\ee
and for anti-bonding states,
\be
\psi_{n',m'} &=& -\phi_{n',m'}=e^{ik_1n' + ik_2m'}, \nonumber \\
E &=& t - (\cos k_1 + \cos k_2),
\ee
where $(n', m')$ are the  coordinates of the square lattice in Fig. \ref{SquareLattice} which are rotated from those in Fig. \ref{HoneyLattice},  and $k_1$ and $k_2$ are the corresponding $k$ vectors which are in region:
\be
-\pi \le k_1 \le \pi, \nonumber \\
-\pi \le k_2 \le \pi.
\label{kspace1}
\ee

Let us apply a magnetic field which gives flux $2\pi \Phi = 2\pi p/q$ per unit cell, where $p$ and $q$ are mutually prime integers. Then each dual space is spilt into $q$ subspaces;
we have $q$ subbands for each sector. In Fig. \ref{HoneyHofst} (d) the lower Hofstadter butterfly is in the bonding sector and the upper Hofstadter butterfly is in the anti-bonding sector.

The quantized values of the Hall conductance for each sector are given as follows \cite{tknn,kohmoto89}: If a Fermi energy is in $r$th gap from the bottom, integer Hall conductance $t_r$ is determined by the Diophantine equation
\be
p t_r+q s_r = r,
\label{diophantine}
\ee
where the integers must satisfy
\be
|t_r| \le \frac{q}{2}, \nonumber \\
0 \le r \le q.
\label{diophantine2}
\ee
The integer Hall conductance carried by $r$th band from the bottom is given by $\sigma_{xy} = 2\pi (t_r - t_{r-1})$ and it is related to the total flux of the dual subspace by the topological relation (\ref{hall}). Thus Hall conductances of subbands in a magnetic field are nonzero as it should be, because time reversal symmetry is broken. 

In Sec \ref{FluxMonopole}, the $2d$ dual space of the honeycomb lattice is extended to $3d$ space $R^3$. The existence of a monoplole-antimonopole pair in $R^3$ is shown in the $SO(3)$ gauge theory. In a magnetic field,  the $2d$ dual space is split into $q$ subspaces by the Hofstadter mechanism. Accordingly $R^3$ is also split into $q$ subspaces $R^3_q$. 
The total Abrikosov flux of a single subband is nonzero since the corresponding quantized Hall conductance is nonzero. This implies that  numbers of monopoles and anti-monopoles are unequal in $R^3_q$. This is {\em monopole-antimonopole  deconfinement due to time reversal symmetry breaking}.

Note that sum of  Hall conductances of all the subbands vanishes. This can be seen from the fact that the only solution of the Diophantaine equation (\ref{diophantine}) with Eq. (\ref{diophantine2}) is $t_r = 0$ for $r= q$. This implies that  the total number of monopoles and antimonopoles are equal in $q$ subspaces $R^3_q$'s. Thus, when a magnetic field is applied,  monopoles and antimonopoles are created in pairs. But they are unpaired and distributed to subspaces $R^3_q$'s.


\section{The Jahn-Teller Effect}
\label{JahnTeller}
The Jahn-Teller effect was first predicted as a very general phenomenon \cite{JT}. It is the intrinsic instability of an electronically degenerate compound against lattice distortions that remove degeneracy. Thus a dynamical lattice symmetry breaking takes place to a lower lattice symmetry. 

As discussed in Sec. \ref{DiracZeroMode}, there are two Dirac modes with $E = 0$ in the electron hopping model on the honeycomb lattice.  Thus if the Fermi energy is at $E = 0$, the ground state is fourfold degenerate. (Each Dirac mode is doubly degenerate.) The condition for the Jan-Teller effect is satisfied but the whole problem is a rather difficult one since all the interactions involoving electrons and nucleus have to be taken into account properly.

Instead of doing it, let us focus on a simpler problem of considering ground state energes of electronic systems with certain periodic modulations of hopping integrals. From Eq. (\ref{KL}) the vector connecting the Fermi surfaces (Dirac points K and L) is given by
\be
{\bm k}_{per3} = (0, \frac{4\pi}{3a}).
\ee
Then we would expect Fermi surface instability modes in M directions with ${\bm k}_{per3}$ whose magnitude is a third of the period of the dual space(see Fig. \ref{MS}). 

\begin{figure}[t]
\begin{center}
\includegraphics[width=30mm]{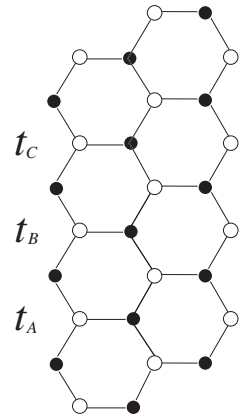}
\caption{The honeycomb lattice with a period $3$ modulation of hopping integrals with $t_A$, $t_B$, and $t_C$ in $y$-direction.
The hopping integrals for the horizontal bonds are  $1$. }
\label{JT0}
\end{center}
\end{figure}

Gap openings for period $3$ modulations in M directions are numerically examined (see Fig. \ref{JT0}). 
Examples are shown in Fig. 11 and we have numerical evidences of  gap openings with period $3$ modulations.
 
\begin{figure}[t]
\begin{center}
\begin{tabular}{c}
\includegraphics[width=55mm]{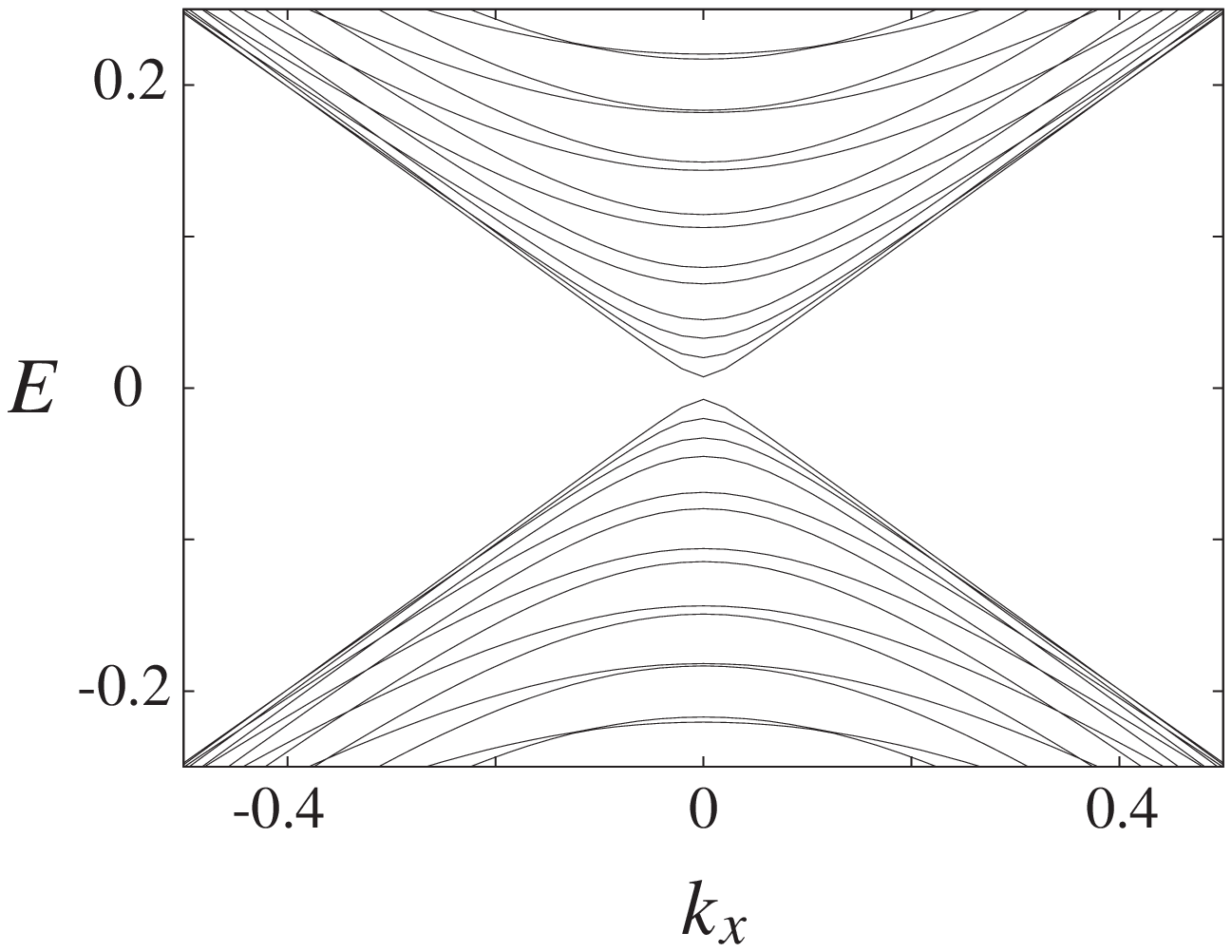}
(a)
\\
\\
\includegraphics[width=55mm]{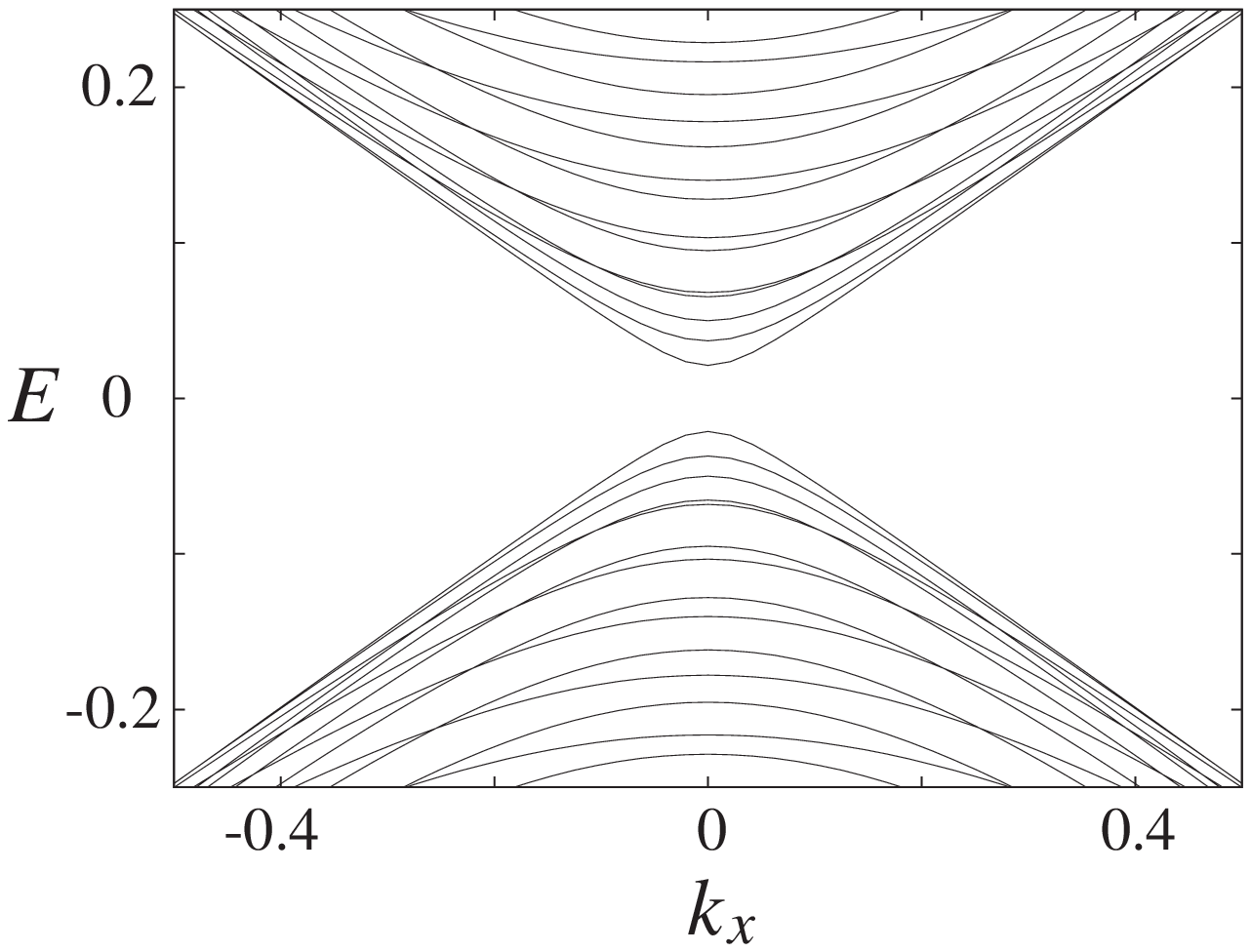}
(b)
\end{tabular}
\caption{Examples of gap openings with period $3$ modulations of hopping integrals. (a) $t_A = t_C = 1.02$ and $t_B = 0.98$. (b) $t_A = t_C = 1.05$ and $t_B = 0.95$.}
\end{center}
\label{JTPeriod3}
\end{figure}
The energy spectra for period $2$ modulations of hopping integrals can be obtained analytically and it is confirmed that there is no gap opening for modulations in neither the  $S^1$ directions nor M directions. Periods $4$ and $5$ modulations are examined numerically and there are no gap openings. From these investigations we conclude that gap openings take place only period $3$ modulations of hopping integrals in M directions.

The situation described above has a certain similarity to the Peirerls instability of a one-dimensional half-filled band in which $2k_F$ instability leads to gap openings and a dimerization of the lattice \cite{peierls}.
Note that density of states for the 1$d$ lattice is finite at $E_F = 0$, but it vanishes for the honeycomb lattice. In this respect the honeycomb lattice is easier to be distorted compared with the 1-$d$ lattices.

In Sec. \ref{1dDirac} the existence of 1$d$ Dirac modes is explicitly demonstrated. 
Therefore  from these modes, we expect to have another types of Jahn-Teller effect if $E_F = \pm t$.

The 1$d$ Dirac modes are on lines $k_y = \pm \pi/2$, and if $E_F = \pm t$, the two Fermi surfaces nest with a nesting vector
\be
{\bm k}_{nest} = ( 0, \frac{2\pi}{a}),
\ee
(it is scaled by Eq. (\ref{rescale})). Thus a possibility of charge density wave(CDW) or spin density wave(SDW) with period $2$ in M direction has to be examined.

\newpage

{\acknowledgements}  
We are especially grateful to J. Friedel for pointing out the existence of Jahn-Teller effect in graphen, and  to T.  Eguchi for informing the author about the work of 't Hooft on magnetic monopoles. We thank D. Tobe for help with computer calculations.



\begin{thebibliography}{99}

\bibitem{novo}
K. S. Novoselov, A. K. Geim, S. V. Morozov, D. Jiang, M. I. Katsnelson, 
I. V. Grigorieva, S. V. Dubonos and A. A. Firsov, 
Nature {\bf 438}, 197 (2005)

\bibitem{zhang}
Y. Zhang, Y.-W Tan, H. Stormer and P. Kim, 
Nature {\bf 438}, 201 (2005).

\bibitem{berger}
C. Berger, Z. Song, X. Li, X. Wu, N. Brown, C. Naud,
D. Mayou, T. Li, J. Hass, A.N. Marchenko, E. H. Conrad,
P. N. First, W. A. de Heer,
Science, \textbf{312}, 1191 (2006).

\bibitem{ann}
M. Kohmoto,
 Ann. of Phys. {\bf 160}, 343 (1985).
 
 \bibitem{FluxUnit}
In physical units, $2\pi$ flux corresponds to the flux quantum $\Phi_0 = hc/e$ (or $\Phi_0 = hc/2e$ for superconductor).

\bibitem{kohmoto89}
 M. Kohmoto,
 Phys. Rev. B{\bf 39}, 11943 (1989).
 
\bibitem{tknn} 
D. Thouless, M. Kohmoto, P. Nightingale, and M. den Nijs,
Phys. Rev. Lett. {\bf 49}, 405 (1982).

\bibitem{HasegawaKohmoto}
Y. Hasegawa and M. Kohmoto,
Phys. Rev. B{\bf 74}, 155415 (2006).

\bibitem{thooft}
G. 't Hooft, Nucl. Phys. B{\bf 79}, 276 (1974).
 
\bibitem{hofst} D. Hofstadter, 
 Phys. Rev. B{\bf 14}, 2239 (1976).
 
\bibitem{JT} H.A. Jahn and E. Teller, Proc. Roy. Soc. {\bf 161}, 220 (1937). 

\bibitem{peierls} R.E. Peierls, {\it Quantum Theory of Solids} (Oxford: Clarendon Press, 1956). 


\end{thebibliography}
\end{document}